%% file: main.tex
\documentclass[journal]{IEEEtran}
\usepackage[utf8]{inputenc}
\usepackage[hidelinks]{hyperref}
\usepackage{multirow}
\usepackage{stfloats}
\usepackage{tikz}
\usetikzlibrary{fit, shadows}

\begin{document}

\title{Behavioural Reports of Multi-Stage Malware}

\author{\IEEEauthorblockN{Marcus~Carpenter, Chunbo~Luo} \\ \IEEEauthorblockA{\textit{Department of Computer Science} \\ \textit{University of Exeter} \\ Exeter, UK \\ Email: mc844@exeter.ac.uk}}

\markboth{IEEE Transactions on Sustainable Computing}%
{Carpenter \MakeLowercase{\textit{et al.}}: Behavioural Reports of Multi-Stage Malware}

\maketitle

\begin{abstract} 
The extensive damage caused by malware requires anti-malware systems to be constantly improved to prevent new threats. The current trend in malware detection is to employ machine learning models to aid in the classification process. We propose a new dataset with the objective of improving current anti-malware systems. The focus of this dataset is to improve host based intrusion detection systems by providing API call sequences for thousands of malware samples executed in Windows 10 virtual machines. A tutorial on how to create and expand this dataset is provided along with a benchmark demonstrating how to use this dataset to classify malware. The data contains long sequences of API calls for each sample, and in order to create models that can be deployed in resource constrained devices, three feature selection methods were tested. The principal innovation, however, lies in the multi-label classification system in which one sequence of APIs can be tagged with multiple labels describing its malicious behaviours.
\end{abstract}

\section{Introduction}
There are billions of malware attacks worldwide every year \cite{StatistaMalware2022}. Most malicious programs are created by cyber-criminals in an attempt to generate profit for organised crime groups. Malware exploit known and unknown vulnerabilities in cyber-systems to gain access into restricted areas of a computer network. Once access is gained, they can steal data, disrupt services, and damage both physical and virtual assets. Anti-malware systems have existed for decades \cite{Denning1987} and are continuously updated to keep their intrusion detection on par with modern attacks. Currently the main trend in the field of intrusion detection is to implement machine learning models that segregate malicious anomalies from benign behaviours \cite{Singh2021}.

Machine learning models require datasets that represent the task to be solved, in this case, detecting malware. The proposed dataset provides the cyber-security community with two main contributions. A textual representation of API calls, and a multi-label classification system for the provided API sequences. This dataset will allow researchers to create and improve host based intrusion detection systems (HIDS). The proposed HIDS monitors the host for API calls and then tag processes with labels that define its behaviour. However, this paper only provides the dataset and a model benchmark as a proof of concept.

This dataset only provides API calls because even though network features are a powerful tool in detecting malware \cite{Wozniak2021}, they inherently involve other hosts. Malicious behaviour can be monitored on the local device because the malware must execute locally on the targeted device. Static features such as packers and imports \cite{Liras2021} were avoided because it is not always possible to analyse a malware file before it executes on a computer. For example, a malicious process that spreads over the network or a fileless malware would still generate API calls, but may leave no trace after the attack. Hardware usage features do not seem to be representative of a malware attack because it is hard to understand how, for example, CPU usage relates to a malware attack even though they have been successfully used to detect them in previous work \cite{Rhode2018}.

The API calls were generated by malware and benign samples running in Windows 10 virtual machines. Windows 10 possesses about 70\% of the market share out of the available Windows operating systems meaning that it still represents a modern computer \cite{OSmarket2022}. Other works using older operating systems may not provide an accurate representation of a modern malware attack. However, it is important to note that the proposed methods and models are not dependant on operating system. The models can be deployed and trained on any device that generates API calls and the multi-label classification system can be applied to any type of malware data, not just API sequences.

The dataset provides API sequences that vary in size from one API call to over seven hundred thousand, and while one API call may not provide much information that aids in classification, overly long sequences contain more words than a novel. It is had to train models on long sequences because of high GPU RAM requirements and slow training times. This constraint is exacerbated if deploying these models on end user devices with limited hardware capabilities such as budget personal computers and IoT devices.

Models need to be updated regularly via retraining to maintain suitable detection performance. This means that the HIDS must contain sustainable models for hardware constrained devices. Therefore, the length of  the API sequences was constricted to an artificial limit of 512 API calls. This number was chosen based on BERT, Google's natural language processing neural network \cite{Devlin2018}. Three feature selection methods where used in order to select the most adequate sub-sequence of APIs, these are explained in Section \ref{sec:Data}.

The main contribution brought by this dataset is the labelling system. To the best of our knowledge all known works so far have focused on mutually exclusive labels. Binary classification can only detect the presence of an intrusion and does not go into detail on what the threat is, making it harder for the analyst to mitigate the attack. Multi-class classification identifies a threat by name and assumes that the analyst will have knowledge on how to mitigate the attack from the given label. The proposed multi-label classification identifies multiple malware types (if present) within the API sequence. This allows the analyst to further understand the malware attack without extensive knowledge of malware taxonomy.

Labelling behaviours and features allows models to detect new malware that does not fit into the current taxonomy because behaviours are common among malware of different classes and families. This means that the multi-label classification system allows new malicious behaviour to be detected even if there is no current specific label for it. In other words, the multi-label classification system is akin to tagging an image with labels such as white, poodle, and toy, instead of classifying it as dog, or poodle.

Section \ref{sec:Design} provides a tutorial on how to built and setup a malware analysis laboratory. Section \ref{sec:Malware-pipeline} provides a tutorial on how to create or expand the dataset. The expandability of the dataset is possible due to its modularity. The data is divided into batches each containing a theme, in this case there is one batch for the benign files and two for malware samples. However, all data is compiled into one when training the models. The dataset is described in detail in Section \ref{sec:Data}. The reason for this modular structure comes from the growing trend of research into concept drift further discussed in Section \ref{sec:conclusion}.

Six different models were trained on each view of the dataset totalling eighteen models in the benchmark. The benchmark results show that the first 512 API calls allow the models to reach higher classification accuracy, while other feature selection algorithms make the data harder to classify. Out of all six model architectures, the GRU performed best on the first feature selection method. However, on the harder slices of the data the bidirectional GRU was the best model arguably making it the superior model overall. The results are further discussed in Section \ref{sec:benchmark}.

The code for the experiments can be found at:

\url{https://github.com/marcusCarpenter97/Malware-data}

\section{Related works}

There has been extensive research on detecting malware using dynamic analysis and API calls to train machine learning models. However, many of these works use features that are undesirable for a HIDS such as network data or use features that may not be necessary for accurate detection of malware. There is also the case for the multi-label classification system proposed by this dataset. Most works focus on binary malware detection and some make use of multi-class classification, but not many provide an opportunity to tag a sample with multiple labels.

There are many datasets that provide a good benchmark for machine learning models. Datasets such as the CTU-13 \cite{Garcia2014}, CICISD2017 \cite{Sharafaldin2018}, and NSL-KDD \cite{Tavallaee2009} are good benchmarks for detecting malware attacks on network data. However, network intrusions are out of scope for this work because the objective is to create a host based detection system and not a network based system. While both, host and network, data can complement each other, they ultimately have different use cases. Network detection focuses on malicious behaviours detectable on IP traffic while the proposed data focuses on detecting malicious behaviours on API calls. It is possible to argue that a HIDS can serve as a backup for network intrusion detection systems were a malware to bypass it, or as a tool to mitigate non-networked transmissions of malware such as USB drives.

Datasets that provide API calls are scarce. Microsoft's Big 2015 dataset \cite{Ronen2018} contains a large amount of malware samples represented as opcodes which allows for natural language processing of the data and the implementation of a HIDS. However, opcodes are extracted from disassembled binary files and it may not always be possible to acquire a malicious file in a live environment, on the other hand monitoring suspicious processes and extracting API calls is a feasible solution for a HIDS. The ADAF-LD dataset \cite{Creech2014} provides API calls for malware running on a Linux environment. The drawbacks of this dataset are that the operating system does not represent a modern environment, and the labels do not support a multi-label classification system.

Aside from the popular datasets mentioned above, many researchers produce their own data or are given data by a private contractor. These datasets tend to be more inline with the proposed dataset, although they are not readily available for the public making them unsuitable for a benchmark. Two types of datasets were reviewed, binary labels and multi-class labels.

Amer et al. \cite{Amer2020} used a Markov Chain for the binary classification of Windows API calls. Word embeddings for each API sequence were clustered into benign and malicious groups with the objective of predicting whether the sequences are malicious or not. The transition probabilities between the two states where calculated using Markov Chains.

Jindal et al. \cite{Jindal2019} classified Cuckoo Sandbox analysis reports as belonging to a malicious or benign file. The JSON reports generated by Cuckoo were processed to fit the fixed standard of the neural network. The model itself used a combination of CNN, LSTM and Attention. Their proposed innovation is that no feature selection is required and the model can choose the best features from the report by itself.

Liras et al. \cite{Liras2021} trained four machine learning binary classifiers, none of them neural networks, on multiple features extracted from dynamic and static analysis of malware. Manual and statistical feature selection was used to extract features that best discriminate each class. The manual method is based on heuristics and knowledge of the domain, and the statistical method involves a small comparison of algorithms that rank features selected by the manual method.

Rhode et al. \cite{Rhode2018} analysed malware with Cuckoo Sandbox, CAPE Sandbox's predecessor, and used a bidirectional GRU to classify the features extracted from the analysis as either malicious or benign. The features selected by the authors were based on the usage of hardware resources. For example, CPU and RAM usage and network packets transmitted. They claim that training models on API calls is hard because APIs are categorical data and unseen APIs cannot be interpolated unless the vocabulary is comprehensive. However, the relation between CPU usage and malicious behaviour is left unexplained.

Imtiaz et al. \cite{Imtiaz2021} trained a neural network to classify android malware. This is one of the few attempts at tagging malware with multiple labels. However, they did not use a multi-label classification system, instead they proposed a two step classification method. First, the sample is classified as either benign or malware and if the file is malicious, it is further labelled into a malware category and a malware family in parallel. Their results show that the models perform better on static features meaning that dynamic analysis provides a harder dataset.

Downing et al. \cite{Downing2021} used an autoencoder model to recreate the binary structure of benign files, and when given a malicious file the model failed to reproduce some sections. These errors were clustered into groups and labelled as a malicious function of the malware, essentially tagging one sample with multiple malicious functionalities. This is different than the proposed multi-label system because it requires the help of an unsupervised clustering algorithm rather than our method that uses a multi-label supervised end-to-end deep learning model.

A few conclusions can be made from this brief literature review. First, binary classification is often preferred even if labelling malware samples with multiple labels at once has been shown to be a powerful tool, likely because binary classification is less challenging. Second, word embeddings have been used to process API calls, although many researchers often combine the APIs with other features rather than use them on their own. And thirdly, feature selection seems to be mostly based on heuristics and domain knowledge although some have attempted to use end-to-end deep learning for feature selection.

\section{System Design}
\label{sec:Design}
Dynamic malware analysis of thousands of samples requires dedicated software running on a powerful workstation. This section explains the choices made when selecting software and hardware components for building the malware analysis laboratory. The software is chosen first as it is the main component of the system while the hardware specifications are defined simply to accommodate the choices in software.

\subsection{Choice in software}
There are four major software components that make up the malware laboratory. The sandbox, the virtual machine, the label maker, and the custom scripts.

The sandbox used to analyse the samples is called CAPE Sandbox \cite{OReilly2022}. CAPE is a project forked from Cuckoo Sandbox and was chosen instead of its predecessor for two reasons. First, Cuckoo is no longer maintained while CAPE is in active development. Second, CAPE has pre-made scripts that help with installation.

The virtual machine (VM) creates a secure environment in which the sandbox can run and analyse the malware. Despite KVM being recommended, VirtualBox was used as the virtualisation software because it is easier to setup. Windows 10 was chosen as the target operating system for the VM because it still holds the majority of the market share \cite{OSmarket2022}.

VirusTotal was used as the label maker. This is a famous and trusted software in the cyber-security community that takes in a file and labels it according to results from over 70 commercial antivirus products. Even though VirusTotal is trusted by the community, Roh et al. state that an automated generation of labels by computer programs results in the weakest type of machine learning label \cite{Roh2021}. However, the alternative of using human experts to label thousands of files would have been too costly in terms of both time and money.

The last software component are the custom scripts for moving and processing the raw data into a usable dataset. Bash scripts and Python were used. These algorithms are presented in the future sections.

\subsection{Choice in hardware}
As mentioned above, hardware requirements exist only to accommodate the software, therefore a different software setup will have different hardware demands.

The sandbox was setup to run 10 VMs in parallel in order to speed up the analysis. Each VM was given 2 cores and 2 GB of RAM, therefore an Intel Core i9-10980XE CPU combined with 64 GB of RAM were used. Storage is also an important component of the system because VMs, thousands of files and their respective reports, plus the final dataset occupy hundreds of GBs, so a 4TB SDD was used. The GPU, while not relevant for the creation of the dataset itself, was used to train the benchmark models. A Nvidia Titan RTX was used as the GPU.

\section{Malware analysis pipeline}
\label{sec:Malware-pipeline}

\begin{figure}
    \centering
    \begin{tikzpicture}
    \tikzset{%
      cascaded/.style = {%
        general shadow = {%
          shadow scale = 1,
          shadow xshift = -1ex,
          shadow yshift = 1ex,
          draw,
          thick,
          fill = white},
        general shadow = {%
          shadow scale = 1,
          shadow xshift = -.5ex,
          shadow yshift = .5ex,
          draw,
          thick,
          fill = white},
        fill = white, 
        draw,
        thick,
        minimum width = 2cm,
        minimum height = 1cm}}
       \node (samples) [cascaded] at (0, 0) {PE samples};

       \node (cape) [rectangle, draw, thick, align=center, minimum height=1.8cm] at (-1, -2)  {C\\A\\P\\E};
       \node (vt) [rectangle, draw, thick, align=center, minimum height=1.8cm] at (1, -2)  {V\\T};

       \draw[->] (samples.south) to (0,-0.8) to (-1,-0.8) to (cape.north);
       \draw[->] (samples.south) to (0,-0.8) to (1,-0.8) to (vt.north);

       \node (comp) [rectangle, draw, thick] at (0,-3.5) {COMPILE};

       \draw[->] (cape.south) to (-1,-3.5) to (comp.west);
       \draw[->] (vt.south) to (1,-3.5) to (comp.east);

        \node (data) at (0,-4.5) {\includegraphics[scale=0.03]{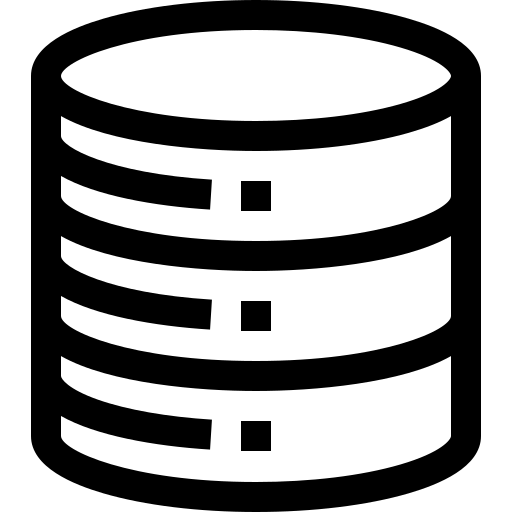}};
        \draw[->] (comp.south) to (data.north);

        \node (datatxt) at (0,-5) {Dataset};
    \end{tikzpicture}
    \caption{System overview of the malware analysis laboratory. The files to be analysed, in this case Windows PE files, are submitted to CAPE Sandbox and VirusTotal (VT) by separate Python scripts. After the analysis is complete, one script extracts and saves the CAPE reports and another performs the same action for the VT reports. From there the compile script processes all results by extracting the relevant information such as API calls and labels and compiles all this data into the multiple batches seen in the final dataset.}
    \label{fig:pipeline}
\end{figure}
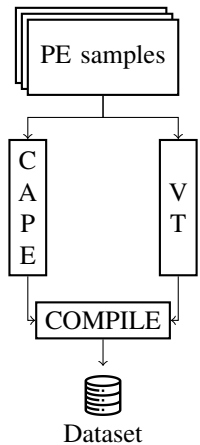

The creation of the proposed dataset can be broken down into four steps of a pipeline: sample collection, labelling malware with VirusTotal, analysing samples with CAPE Sandbox, and processing the collected data. The analysis and the labelling can be completed in parallel to save time. Each step is described below and its accompanied by a diagram detailing the whole pipeline in Figure \ref{fig:pipeline}.

\subsection{Sample collection}
Samples were downloaded from the VirusShare repository \footnote{https://virusshare.com/}. The data from VirusShare comes in batches of tens or hundreds of gigabytes in size which are downloaded via torrent. However, not all samples in a batch are desirable. Only the files that contained a portable executable (PE) header were selected for this dataset. These are Windows executable files (exe or dll).

To remove undesirable samples the Python module `pefile' \footnote{https://pypi.org/project/pefile/} was used in a custom script. If the PE header is not detected in a file by the third party module, then the file is deleted. This rule was applied to all files downloaded from VirusShare. However, collecting samples is more than just acquiring malware.

Benign files are also part of the dataset albeit they only compose a small section of it. This task was more complicated than acquiring malware samples. It is of course possible to use system files from the Windows operating system, and some were used, but manually downloading benign apps is a tedious chore without a web crawler. And because the program must run during the analysis, these benign apps must either already be installed in the system or the installer itself must be analysed.

\subsection{Labelling malware with VirusTotal}
VirusTotal provides a web service that receives files, analyses them using their custom setup, and produces a downloadable report. Once an account is registered (with an academic API), the API can be accessed through various means, all of which can be found in their official GitHub repository \footnote{https://github.com/VirusTotal/vt-py}.

Only the malicious files were submitted to VirusTotal because benign files already have a label by default, although this could generate interesting results in terms of false positives and the trust placed in companies to provide non-malicious software.

Custom Python scripts were written to handle the submission of files and the retrieval of reports from VirusTotal. The reports provide a multi-label classification for each file based on over 70 commercial antivirus products. This means that one malware sample can have one or more labels defining its class. The label structure will be further discussed in Section \ref{sec:Data}.

\subsection{Analysing samples with CAPE Sandbox}
CAPE Sandbox has multiple components and requires careful configuration for all of them to work. This section briefly explains how everything is setup and then how the analysis process works. Once setup, the analysis can occur in parallel to the VirusTotal labelling stage.

CAPE can be installed by running the helper script provided in the official GitHub page \cite{OReilly2022}. Once installed there are two parts to consider when configuring CAPE Sandbox, configuration files and virtual machines.

There are two configuration files that need to be edited before CAPE can work properly, `cuckoo.conf' and `machinery.conf', both detailed in the documentation. In the `cuckoo.conf' file the `machinery' field was edited to use VirtualBox because KVM is the default choice. The only other field that needs changing is the IP address under the ResultServer section which should take the IP address of the hypervisor's network interface.

The `machinery.conf' file takes the name of the hypervisor of choice, in this case `virtualbox.conf'. The file structure is comprehensive but only contains one entry. If using multiple VMs, simply copy and paste the configuration under a different name and IP address within the same file. The VMs must be setup once CAPE is configured. The VM's name must match whatever was inserted in the `machinery.conf' file. Basic software is required to analyse certain files for example if analysing .docx, MS Word must be installed.

The agent is a script that allows the VM to communicate with the sandbox during the analysis of a file. The agent must be executed with administrator privileges inside the VM. It is also important to deactivate the firewall and any Windows 10 feature that blocks the execution of potentially malicious software. This includes Windows Defender, SmartScreen, and User Account Control, etc.

The IP of the VM must be fixed for the sandbox to work, this can be achieved via Windows networking configuration and the virtualisation software of choice must be set to host-only networking. The VM must be able to ping the host and vice-versa. A snapshot of the fully configured VM running the agent script is advised, as CAPE can be configured to directly start the VM from this state.

Once the VM and CAPE are fully configured and working the analysis can begin. A custom Bash script was written to submit samples for analysis. The script uses CAPE's submission utility to send samples to each of the ten VMs in an alternating manner. This guarantees that each VM will execute in parallel.

\subsection{Processing collected data}
After the CAPE and VirusTotal analyses are complete, all result files must be processed into a usable dataset. There are two algorithms that were used to process the two parts of the data, one to extract API calls from CAPE reports, and another to transform the VirusTotal reports into labels for machine learning models.

The reports generated by CAPE Sandbox are JSON files and the API calls are stored under the following set of keys: \verb|behavior|, \verb|processes|, \verb|calls| and then there are the \verb|timestamp| and \verb|api| keys. Each malware can spawn one or more processes, and each process generates a sequence of timestamped APIs. The APIs along with their timestamps are extracted from the JSON report into one sequence and sorted by time.

Sub-processes were not considered separately because models can, or ideally should, detect relationships between non-consecutive APIs. Moreover, as the API calls are sorted by time the malicious events can still be modelled in the order in which they occurred in the VM.

Similar to the CAPE reports, VirusTotal reports are also stored in JSON format and the labels used are under the following set of keys: \verb|data|, \verb|attributes|, \verb|popular_threat_classification|, \verb|popular_threat_category|. This contains textual labels describing potential malware classes for a sample. These textual labels are converted into a binary array containing fifteen digits based on a lookup table generated from all unique textual labels in the dataset. Each position in the binary array represents a class in the data, including benign and generic malware that cannot be further classified. Each set bit in the binary array indicates that the API sequence belongs to a specific class in the data.

\section{The data}
\label{sec:Data}

There are three points to explain about the proposed dataset. First, this section explains the file structure of the dataset and the data structure in which the dataset is formatted in. The second point is a discussion on how feature selection was applied onto the API sequences to reduce hardware resource usage and optimise training times. And the third and final point is on the label structure and the label distribution for the training and testing sets.

The data has a physical and logical structure. The physical structure details how the files are organised, and the logical structure relates to the internal data structure of the file once it is loaded into memory. The data is composed of JSON files, each file representing a batch of samples. A batch is a logical partition of data that contains four fields, name, year, APIs, and labels. The name is an arbitrary label given to the batch, in this particular dataset most batch names relate to VirusShare names. The year is the year in which the samples were collected, also extracted form VirusShare.

The APIs are stored as a list of strings, each string containing one function call name. Some samples generate multiple processes during their execution, these multiple API branches were merged into one sequence according to their timestamp. Therefore all API calls are represented chronologically but information about child and parent processes is lost. The data provides very long API sequences that make training models very challenging. The length of the API sequences varies greatly from one system call to over seven hundred thousand, the latter is seven times longer than the word count of the average novel.

Training neural networks on long textual sequences is only feasible in small models but not scalable for two main reasons, one the GPU RAM is a limiting factor, and two the training times become increasingly longer. To address the variable lengths of the API sequences, only 512 API calls per sample were processed by the model, this number is based on Google's BERT model \cite{Devlin2018}. Using only 512 API calls allowed for models to be trained faster and with a reasonable GPU RAM usage. If only 512 API calls are to be processed by the model, then the selected APIs must contain the greatest amount of information that can be used to classify the sample. For this reason three feature selection experiments were conducted.

Each feature selection method created a different view of the same data. All views have the same train and test label distribution. The first view only uses the first 512 API calls from the sequence, this is called `First'. The second view uses the last 512 API calls, this is called `Last'. The third takes a random slice of 512 consecutive APIs from the sequence, this is called `Random'. In each of the three views, if any API sequence is shorter than 512 API calls, then the sequence is used as a whole and padded later on.

The labels are stored as binary arrays to accommodate the multi-labelling classification system. Each label is one row of 15 columns, and each column represents one possible class the sample can be labelled as. The values in each cell can only be 0 or 1, where 0 represents a lack of membership for the class related to the column and 1 represents membership. A sample must belong to at least one class, and in theory can belong to all classes at once. However, in practice there are some mutually exclusive labels. For example, benign labels are not marked as malicious, or malware without any further classification can only be labelled as generic malware. In the future work, Section \ref{sec:conclusion}, the possibility of labelling benign files with malicious classes is discussed.

To make experiments reproducible the Python 3.9 module `random' used the seed 888 and the \verb|train_test_split| method from sklearn 1.0.2 used the random state parameter at 888. Table \ref{tab:labeldistribution} shows the result of the train and test split along with the distribution of samples for each available label. The main observation to make is that the classes in the data are heavily imbalanced and some might not have enough samples for a machine learning model to learn from. There are 8087 samples in the dataset, although label count is greater because of the multi-label system.

\begin{table}
    \centering
    \caption{Distribution of labels in dataset}
    \input{Tables/label-distribution.tex}
    \label{tab:labeldistribution}
\end{table}

\section{The models}
\label{sec:models}

This section explains the six model architectures used to produce the benchmark. Each model architecture was trained on all three views created by the feature selection making a total of eighteen models. The six architectures were the Simple Embedding Model (SEM), Multilayer Perceptron (MLP), Long-Short Term Memory (LSTM), Gated Recurrent Unit (GRU), and the bidirectional versions of the LSTM and GRU.

All six models used the Adam algorithm at \(0.001\) learning rate to optimise the binary cross entropy loss function. Fifteen epochs were used to train the models, although early stopping was employed to prevent unnecessary iterations.

The most basic model is the Simple Embedding Model (SEM). First, each sequence of APIs is converted into a string and then vectorized. Vectorization is the process of converting text into a sequence of integers, where each unique token is mapped onto a unique integer. The vectorization layer is only trained on the training corpus, and not the testing data, as a means to prevent data leakage.

The second step of the SEM is to train an embedding layer on the vectorized input. This helps the model learn how to best represent each API call in the sequence and the relations between them by assigning each system call a set of floating point features. The experiments in this paper used an embedding dimension of 16.

The third and final step of the SEM is the output layer which converts the learnt embeddings into a multi-label binary array. The output is produced by a fully connected layer with the same number of nodes as classes in the label. The activation function is the Sigmoid. The SEM is a very simple model, text vectorization, embedding, and output, that serves as a foundation for the other five models.

Both the SEM and the MLP require a Global Average Pooling layer between the embedding layer and the fully connected layer to process the multi-dimensional data. All models, except the SEM, use a fully connected layer with 64 nodes and a ReLU activation before the output layer. The recurrent models build upon the MLP by replacing the Global Average Pooling layer with one of the following layers: LSTM, GRU, and the bidirectional version of each. All recurrent layers also have 64 nodes.

\begin{table}[t]
    \centering
    \caption{Comparison of accuracy and F1-score}
    \input{Tables/acc-f1-table.tex}
    \label{tab:acc-f1-table}
\end{table}

\section{Benchmark results}
\label{sec:benchmark}

The benchmark contains two types of results based on six error metrics which are defined in sub-section \ref{sec:error-metrics}. The first set of results provides an overall model benchmark for all experiments, this is detailed in sub-section \ref{sec:model-benchmark}. The second set of results explains how the model performed for each malware class in the data, this is discussed in sub-section \ref{sec:performance-malclass}.

Each table separates the error metrics by feature selection method. The `First' only used the first 512 APIs from the entire sequence. The `Last' only used the last 512 APIs from the sequence. And the `Random' took a random slice of 512 consecutive APIs from the whole sequence.

\subsection{Error metrics}
\label{sec:error-metrics}

Two famous machine learning Python packages were used to calculate the error values for the models. Scikit learn and Tensorflow. And six metrics were used to calculate the errors. The accuracy, F1-score, precision, recall, and the binary cross entropy which was used as the loss function for optimising the models. There are only five metrics in the list because Scikit learn and Tensorflow provide different accuracy measurements.

The Scikit learn accuracy only provides a score if the predicted label matches the ground truth in its entirety. On the other hand, the binary accuracy provided by Tensorflow considers each individual element in the binary array when scoring the output of the model. This means that the Scikit accuracy is more punitive than Tensorflow's.

Scikit learn computes the F1-score as: \[F1 = 2 \times (precision \times recall) \div (precision + recall)\]

This metric was used in two ways. First by averaging the F1 over all samples using the \verb|average=samples| parameter to provide a global score for each model, results can be found in Table \ref{tab:acc-f1-table}. Second was by calculating the F1 scores for each class in the data with \verb|average=None|, these results can be found in Tables \ref{tab:first-f1-table}, \ref{tab:last-f1-table}, and \ref{tab:random-f1-table}. Both methods used the parameter \verb|zero_division=1|.

Table \ref{tab:tensorflow-eval} provides the results from Tensorflow's  built in evaluate method. This differs from the other tables where the predictions and labels were used by Scikit learn to calculate error values. The loss is the binary cross entropy and the other metrics include the binary accuracy, explained above, precision and recall. The loss is the only metric that should be minimised, all other values are meant to be maximised. In other words the higher the better.

\subsection{Model benchmark}
\label{sec:model-benchmark}

The model benchmark is split between two tables. Table \ref{tab:acc-f1-table} details the results calculated by Scikit learn by using the labels and the model's predictions. And, Table \ref{tab:tensorflow-eval} contains the results generated by Tensorflow's evaluate function.

Table \ref{tab:acc-f1-table} shows that the GRU was the best performing model on the `First' data slice, and the BiGRU achieved best performance on the `Last' and `Random' slices. The `First' data slice is easier to classify than the `Last' and `Random', this finding is based on the error values observed on the tables. This phenomenon shows that the malware's behaviour is more distinguishable in the first 512 API calls than in the rest of the sequence, likely due to it containing the launch of the attack.

The GRU outperforms non-sequential models because its recurrence is ideal for the sequential nature of the API data, and in this case they outperformed the LSTM model even though Chung et al. found no difference between them other than reduced computational cost \cite{Chung2014}. The bidirectional GRU outperforms the GRU because of its extra backwards layer that allows it to find meaningful connections between the API calls that have occurred and future calls adding context to the current sequence.

Table \ref{tab:tensorflow-eval} corroborates previous results further proving that the GRU is the best performing model in the `First' data slice, and the BiGRU is the best model overall. Scikit Learn's accuracy score illustrates that the models are randomly guessing the labels because of the scores around 0.5. However, Tensorflow's accuracy reaches over 0.9 because they match each digit in the binary array separately inflating the accuracy by counting all the predicted 0s. This is only possible because of the sparsity of the labels.

The precision in Table \ref{tab:tensorflow-eval} is of about 0.8 meaning that the models do not over predict the labels, in other words they do not produce too many false positives when detecting malware. However, this is the only metric in which the BiGRU does not outperform the other models. The recall, however, is around 0.6, slightly above random guessing meaning that there are many samples which the models cannot properly tag as malware, likely due to there not being many samples in each class for the model to learn from. Overall the models achieve an F1 score between 0.6 and 0.7 (Table \ref{tab:acc-f1-table}) making the benchmark good while also allowing for further improvement.

\subsection{Performance on malware classes}
\label{sec:performance-malclass}

\begin{table*}
    \centering
    \caption{Comparison of binary cross entropy (loss) and three other error metrics on each data slice of the test set}
    \resizebox{\textwidth}{!}{%
    \input{Tables/eval-metrics.tex}}
    \label{tab:tensorflow-eval}
\end{table*}

Tables \ref{tab:first-f1-table}, \ref{tab:last-f1-table} and \ref{tab:random-f1-table} detail the model's performance on the individual classes in each of the data slices. The results show that the `First' data slice produces better trained models than the `Last' and the `Random' data slices, in that order. Better trained models are defined as having higher scores in their classification metrics.

From the tables mentioned above, it is possible to make four conclusions on how the models detect each class in the data. Even though the classes are not mutually exclusive, these conclusions are made on each individual class.

Benign, Dropper, Spyware, Miner, Hacktool, and Fakeav, are the hardest classes to detect. Only the BiGRU on the First data slice detected benign samples with a low F1 score of 0.21. It is possible that this is a statistical outlier and the model did not learn how to classify the benign samples. The other classes were not detected by any of the models in any of the data slices. This is likely because there are not enough training samples in each of these categories with the biggest containing 236 samples.

Pua and Downloader have 544 and 452 samples respectively and were detected more often by the models in the `First' slice while not detected in the `Last' and `Random' slices. The Adware class contains 642 samples in the train set, a quantity big enough to train the models consistently at a reasonable F1 score on the `First' slice. This performance decreased on the `Last' slice and became more unstable on the `Random' slice.

Unlike other classes, the Ransomware, Worm, and Virus, are detected by many models in the `First' slice with good F1 scores even though they have a small number of samples. The largest class size is 208. Models performing better on classes with less samples implies that the malware types are easier to classify due to some feature in the API sequences that compensates for the lack of data.

The most populous classes are the Trojan and the Banker malware, followed by the generic malware class. A label reserved for when there are no other labels available but the file is known to be malicious. The size of these classes makes it easier for the neural networks to learn how to classify them with good F1 scores. However, the generic malware class scores low on the F1 score in all three data slices even though it contains 905 samples in the training set.

The difficulty in classifying the generic malware class comes from the fact that these API sequences may actually belong to other classes but labels were missing. The variance in the data creates a challenge for the neural networks to reach adequate performance.

\section{Conclusion}
\label{sec:conclusion}

The proposed dataset aims to address the lack of research on classifying malware based on dynamic analysis. The data introduces malware attacks represented as a textual sequence of API calls. In addition, the main innovation is the multi-label classification system that allows one malware attack to be tagged with multiple malware classes. This new classification system was benchmarked using six neural network models and three feature selection methods. Moreover, a tutorial on how to create a malware analysis laboratory and how to compile the dataset was given.

Five potential future work directions include: addressing the class imbalance of the dataset by either improving the models or expanding the dataset by collecting more samples; some benign software can be labelled as malicious either as a false positive or as a potentially malicious behaviour, these samples could be added to the data in an attempt to determine how much trust can be placed on software development companies; modifying the labelling system to include the probabilities of a sample belonging to a class rather than a binary membership value, this would rely on a consensus of how many antivirus label a sample as a certain class; sub-classing malware into families, this was avoided because it would generate an overwhelming amount of labels that may not be significant; studying concept drift and creating models that can accurately classify malware across multiple time periods.

\begin{table*}
    \centering
    \caption{F1 score on each class of first data slice of the test set}
    \resizebox{\textwidth}{!}{%
    \input{Tables/first-f1-table.tex}}
    \label{tab:first-f1-table}
\end{table*}

\begin{table*}
    \centering
    \caption{F1 score on each class of last data slice of the test set}
    \resizebox{\textwidth}{!}{%
    \input{Tables/last-f1-table.tex}}
    \label{tab:last-f1-table}
\end{table*}

\begin{table*}
    \centering
    \caption{F1 score on each class of random data slice of the test set}
    \resizebox{\textwidth}{!}{%
    \input{Tables/random-f1-table.tex}}
    \label{tab:random-f1-table}
\end{table*}

\bibliographystyle{IEEEtran}
\bibliography{main}

\end{document}

%% file: Tables/label-distribution.tex
\begin{tabular}{lrrr}
\hline
 Name       &   Train &   Test &   Total \\
\hline
 benign     &      66 &     35 &     101 \\
 malware    &     905 &    426 &    1331 \\
 trojan     &    4369 &   2167 &    6536 \\
 banker     &    1089 &    521 &    1610 \\
 pua        &     544 &    278 &     822 \\
 downloader &     452 &    266 &     718 \\
 adware     &     642 &    332 &     974 \\
 dropper    &     236 &    108 &     344 \\
 spyware    &     122 &     65 &     187 \\
 virus      &     208 &    118 &     326 \\
 miner      &      69 &     33 &     102 \\
 ransomware &     177 &     87 &     264 \\
 worm       &     155 &     82 &     237 \\
 hacktool   &      87 &     36 &     123 \\
 fakeav     &       4 &      6 &      10 \\
\hline
\end{tabular}

%% file: Tables/acc-f1-table.tex
\begin{tabular}{lllllll}
\hline
\multirow{2}{*}{Models} & \multicolumn{2}{c}{First} & \multicolumn{2}{c}{Last} & \multicolumn{2}{c}{Random} \\
\cline{2-7}
    & Acc & F1 & Acc & F1 & Acc & F1 \\
\hline
SEM     & 0.443 & 0.682 & 0.407 & 0.655 & 0.414 & 0.655 \\
MLP     & 0.510 & 0.711 & 0.467 & 0.690 & 0.426 & 0.662 \\
LSTM    & 0.416 & 0.656 & 0.448 & 0.674 & 0.408 & 0.654 \\
GRU     & \textbf{0.531} & \textbf{0.722} & 0.479 & 0.687 & 0.452 & 0.673 \\
BiLSTM  & 0.487 & 0.684 & 0.480 & 0.695 & 0.441 & 0.665 \\
BiGRU   & 0.525 & 0.707 & \textbf{0.483} & \textbf{0.698} & \textbf{0.467} & \textbf{0.685} \\
\hline
\end{tabular}

%% file: Tables/eval-metrics.tex
\begin{tabular}{lllllllllllll}
\hline
\multirow{2}{*}{Models} & \multicolumn{3}{l}{Loss} & \multicolumn{3}{l}{Binary accuracy} & \multicolumn{3}{l}{Precision} & \multicolumn{3}{l}{Recall} \\
\cline{2-13}
    & First & Last & Random & First & Last & Random & First & Last & Random & First & Last & Random \\
\hline
SEM     & 0.177 & 0.184 & 0.19  & 0.94  & 0.937 & 0.937 & 0.845 & 0.824 & 0.832 & 0.581 & 0.575 & 0.565 \\
MLP     & 0.159 & 0.172 & 0.183 & 0.946 & 0.943 & 0.938 & 0.847 & 0.847 & 0.84  & 0.649 & 0.611 & 0.569 \\
LSTM    & 0.181 & 0.18  & 0.198 & 0.938 & 0.941 & 0.937 & 0.823 & \textbf{0.85} & 0.821 & 0.585 & 0.594 & 0.571 \\
GRU     & \textbf{0.148} & 0.171 & 0.178 & \textbf{0.951} & 0.942 & \textbf{0.942} & \textbf{0.87} & 0.836 & \textbf{0.86} & \textbf{0.671} & 0.616 & 0.586 \\
BiLSTM  & 0.165 & 0.171 & 0.189 & 0.945 & 0.943 & 0.938 & 0.858 & 0.843 & 0.827 & 0.623 & 0.621 & 0.581 \\
BiGRU   & 0.149 & \textbf{0.168} & \textbf{0.17} & 0.949 & \textbf{0.944} & \textbf{0.942} & 0.856 & 0.849 & 0.844 & 0.664 & \textbf{0.626} & \textbf{0.61} \\
\hline
\end{tabular}

%% file: Tables/first-f1-table.tex
\begin{tabular}{llllllllllllllll}
\hline
 Model   &   benign &    malware &   trojan &   banker &       pua &   downloader &   adware &   dropper &   spyware &    virus &   miner &   ransomware &     worm &   hacktool &   fakeav \\
\hline
 SEM     & 0     & 0.399 & 0.909 & 0.824 & 0     & 0     & 0.011 & 0 & 0 & 0     & 0 & 0.242 & 0     & 0 & 0 \\
 MLP     & 0     & 0.385 & 0.908 & 0.874 & 0.069 & 0     & 0.543 & 0 & 0 & 0.613 & 0 & 0.562 & 0.728 & 0 & 0 \\
 LSTM    & 0     & 0.004 & 0.897 & 0.83  & 0     & 0     & 0     & 0 & 0 & 0.526 & 0 & 0     & 0.656 & 0 & 0 \\
 GRU     & 0     & 0.462 & 0.912 & 0.9   & 0.227 & 0.132 & 0.691 & 0 & 0 & 0.587 & 0 & 0.549 & 0.696 & 0 & 0 \\
 BiLSTM  & 0     & 0.401 & 0.9   & 0.859 & 0.021 & 0     & 0.593 & 0 & 0 & 0.537 & 0 & 0.585 & 0.691 & 0 & 0 \\
 BiGRU   & 0.21  & 0.378 & 0.906 & 0.899 & 0.238 & 0.195 & 0.678 & 0 & 0 & 0.61  & 0 & 0.592 & 0.748 & 0 & 0 \\
\hline
\end{tabular}

%% file: Tables/last-f1-table.tex
\begin{tabular}{llllllllllllllll}
\hline
 Model   &   benign &   malware &   trojan &   banker &   pua &   downloader &   adware &   dropper &   spyware &    virus &   miner &   ransomware &      worm &   hacktool &   fakeav \\
\hline
 SEM     &        0 & 0.022 & 0.895 & 0.864 &     0 &    0     & 0     &         0 &         0 & 0.287 &       0 &    0     & 0.07  &          0 &        0 \\
 MLP     &        0 & 0.401 & 0.907 & 0.87  &     0 &    0     & 0.117 &         0 &         0 & 0.581 &       0 &    0     & 0.745 &          0 &        0 \\
 LSTM    &        0 & 0.307 & 0.901 & 0.859 &     0 &    0     & 0.116 &         0 &         0 & 0.545 &       0 &    0     & 0.689 &          0 &        0 \\
 GRU     &        0 & 0.384 & 0.901 & 0.877 &     0 &    0     & 0.388 &         0 &         0 & 0.548 &       0 &    0.022 & 0.693 &          0 &        0 \\
 BiLSTM  &        0 & 0.352 & 0.907 & 0.882 &     0 &    0     & 0.337 &         0 &         0 & 0.554 &       0 &    0     & 0.703 &          0 &        0 \\
 BiGRU   &        0 & 0.351 & 0.909 & 0.883 &     0 &    0.064 & 0.364 &         0 &         0 & 0.563 &       0 &    0.085 & 0.698 &          0 &        0 \\
\hline
\end{tabular}

%% file: Tables/random-f1-table.tex
\begin{tabular}{llllllllllllllll}
\hline
 Model   &   benign &   malware &   trojan &   banker &       pua &   downloader &   adware &   dropper &   spyware &     virus &   miner &   ransomware &     worm &   hacktool &   fakeav \\
\hline
 SEM     &        0 &  0.2   & 0.897 & 0.829 & 0     &            0 & 0     &         0 &         0 & 0     &       0 &            0 & 0     &          0 &        0 \\
 MLP     &        0 &  0.305 & 0.9   & 0.832 & 0     &            0 & 0     &         0 &         0 & 0     &       0 &            0 & 0     &          0 &        0 \\
 LSTM    &        0 &  0     & 0.896 & 0.856 & 0     &            0 & 0     &         0 &         0 & 0     &       0 &            0 & 0     &          0 &        0 \\
 GRU     &        0 &  0.388 & 0.905 & 0.853 & 0     &            0 & 0.303 &         0 &         0 & 0.065 &       0 &            0 & 0     &          0 &        0 \\
 BiLSTM  &        0 &  0.302 & 0.9   & 0.814 & 0     &            0 & 0.275 &         0 &         0 & 0     &       0 &            0 & 0     &          0 &        0 \\
 BiGRU   &        0 &  0.381 & 0.905 & 0.88  & 0.06  &            0 & 0.394 &         0 &         0 & 0.419 &       0 &            0 & 0.442 &          0 &        0 \\
\hline
\end{tabular}